\documentclass[iop]{emulateapj}
\usepackage[usenames,dvipsnames]{xcolor}
\usepackage{multirow}

\newcommand\cii{[C~{\sc ii}]~}
\newcommand\oiii{[O~{\sc iii}]~}
\newcommand\nii{[N~{\sc ii}]~}

\shorttitle{CNO emission of an unlensed submillimeter galaxy}
\shortauthors{Tadaki et al.}

\begin{document}


\title{CNO emission of an unlensed submillimeter galaxy at $z=4.3$}


\author{Ken-ichi Tadaki\altaffilmark{1},
Daisuke Iono\altaffilmark{1,2},
Bunyo Hatsukade\altaffilmark{3},
Kotaro Kohno\altaffilmark{3,4},
Minju M. Lee\altaffilmark{5},
Yuichi Matsuda\altaffilmark{1,2},
Tomonari Michiyama\altaffilmark{1,2},
Kouichiro Nakanishi\altaffilmark{1,2},
Tohru Nagao\altaffilmark{6},
Toshiki Saito\altaffilmark{7},
Yoichi Tamura\altaffilmark{8},
Junko Ueda\altaffilmark{1},
Hideki Umehata\altaffilmark{9}
}


\affil{\altaffilmark{1} National Astronomical Observatory of Japan, 2-21-1 Osawa, Mitaka, Tokyo 181-8588, Japan; tadaki.ken@nao.ac.jp}
\affil{\altaffilmark{2} Department of Astronomical Science, SOKENDAI (The Graduate University for Advanced Studies), Mitaka, Tokyo 181-8588, Japan}
\affil{\altaffilmark{3} Institute of Astronomy, Graduate School of Science, The University of Tokyo, 2-21-1 Osawa, Mitaka, Tokyo 181-0015, Japan}
\affil{\altaffilmark{4} Research Center for the Early Universe, Graduate School of Science, The University of Tokyo, 7-3-1 Hongo, Bunkyo-ku, Tokyo 113-0033, Japan}
\affil{\altaffilmark{5} Max-Planck-Institut f\"ur extraterrestrische Physik (MPE), Giessenbachstr., D-85748 Garching, Germany}
\affil{\altaffilmark{6} Research Center for Space and Cosmic Evolution, Ehime University, Matsuyama, Ehime 790-8577, Japan}
\affil{\altaffilmark{7} Max-Planck Institute for Astronomy, K\"onigstuhl, 17 D-69117 Heidelberg, Germany}
\affil{\altaffilmark{8} Division of Particle and Astrophysical Science, Nagoya University, Furocho, Chikusa, Nagoya 464-8602, Japan}
\affil{\altaffilmark{9} The Institute of Physical and Chemical Research (RIKEN), 2-1 Hirosawa, Wako-shi, Saitama 351-0198,
Japan}


\begin{abstract}
We present the results from ALMA observations of \nii205 $\mu$m, \cii158 $\mu$m, and \oiii88 $\mu$m lines in an unlensed submillimeter galaxy at $z=4.3$, COSMOS-AzTEC-1, hosting a compact starburst core with an effective radius of $\sim$1 kpc.
The \cii and \nii emission are spatially-resolved in 0.3\arcsec-resolution (1 kpc in radius).
The kinematic properties of the \nii emission are consistent with those of the CO(4-3) and \cii emission, suggesting that the ionized gas feels the same gravitational potential as the associated molecular gas and photodissociation regions (PDRs).
On the other hand, the spatial extent is different among the lines and dust continuum: the \cii emitting gas is the most extended and the dust is the most compact, leading to a difference of the physical conditions in the interstellar medium.
We derive the incident far-ultraviolet flux and the hydrogen gas density through PDR modeling by properly subtracting the contribution of ionized gas to the total \cii emission.
The observed \cii emission is likely produced by dense PDRs with $n_\mathrm{H}^\mathrm{PDR}=10^{5.5-5.75}$ cm$^{-3}$ and $G_0=10^{3.5-3.75}$ in the central 1 kpc region and $n_\mathrm{H}^\mathrm{PDR}=10^{5.0-5.25}$ cm$^{-3}$ and $G_0=10^{3.25-3.5}$ in the central 3 kpc region.
We have also successfully measured the line ratio of [O~{\sc iii}]/[N~{\sc ii}] in the central 3 kpc region of COSMOS-AzTEC-1 at $z=4.3$, which is the highest redshift where both nitrogen and oxygen lines are detected. 
Under the most likely physical conditions, the measured luminosity ratio of $L_\mathrm{[OIII]}/L_\mathrm{[NII]}=6.4\pm2.2$ indicates a near solar metallicity with $Z_\mathrm{gas}=0.7-1.0~Z_\odot$, suggesting a chemically evolved system at $z=4.3$.

\end{abstract}





\section{Introduction}

\vspace{5mm}

\noindent
Submillimeter bright galaxies (SMGs) at $z>3$ are the most likely progenitors of elliptical galaxies in the present-day Universe. 
They are intensively forming stars in the central 1 kpc region \citep[e.g.,][]{2015ApJ...810..133I,2015ApJ...798L..18H} and massive with a stellar mass of $M_\star>10^{11}~M_\odot$ \citep{2014A&A...571A..75M}.
The size is comparable to massive, compact quiescent galaxies at $z\sim2$, which could eventually evolve into larger ellipticals by dry mergers \citep[e.g.,][]{2015ApJ...813...23V}.
These results suggest an evolutionary link from SMGs at $z>3$ to ellipticals at $z=0$.

\begin{table*}[!t]
\caption{Summary of observations and line properties in AzTEC-1. \label{tab;obs}}
\begin{center}
\begin{tabular}{lcccccc}
\hline
 & & ALMA Band-3 & ALMA Band-6 & ALMA Band-7 & ALMA Band-9&\\
 & & CO (4-3) & \nii205 $\mu$m & \cii158 $\mu$m & \oiii88 $\mu$m  \\
\hline
\multicolumn{6}{c}{observations and imaging parameters} \\
\hline
observation date & & 2017/10,11 & 2018/11 & 2017/12 & 2018/11,12\\
baseline length & (k$\lambda$) & 12--4664 & 14--1277 & 18--2967 & 32--2017 \\
\multirow{2}{*}{frequency coverage} & \multirow{2}{*}{(GHz)} & 85.4--89.1    & 256.0--259.8 & 342.8--346.7 & \multirow{2}{*}{630.5--638.0} \\
                                                           &                                    & 97.5--101.2 & 270.9--274.7 & 354.9--258.7 & \\
on-source time & (min) & 420 & 300 & 32 & 100 \\
\multicolumn{2}{l}{{\it uv} tapering for $0.3^{\prime\prime}$-resolution maps} & 0.2$^{\prime\prime}$ & n/a & 0.2$^{\prime\prime}$ & 0.1$^{\prime\prime}$ \\
\multicolumn{2}{l}{{\it uv} tapering for $0.9^{\prime\prime}$-resolution maps}& 0.6$^{\prime\prime}$ & 0.4$^{\prime\prime}$ & 0.5$^{\prime\prime}$ & 0.5$^{\prime\prime}$ \\
$1\sigma_{50\mathrm{km}, 0.3^{\prime\prime}}$ & (mJy beam$^{-1}$)  & 0.10 & 0.09 & 0.41 & 1.5 \\
$1\sigma_{50\mathrm{km}, 0.9^{\prime\prime}}$ & (mJy beam$^{-1}$)  & 0.25 & 0.17 & 0.89 & 3.0 \\
\hline
\multicolumn{6}{c}{fluxes and luminosities} \\
\hline
$S_\mathrm{peak, 0.3^{\prime\prime}}dv$ & (Jy beam$^{-1}$ km s$^{-1}$) & 0.53$\pm$0.02 & 0.35$\pm$0.02 & 4.23$\pm$0.08 & $<$0.88 &  \\
$S_\mathrm{peak, 0.9^{\prime\prime}}dv$ & (Jy beam$^{-1}$ km s$^{-1}$) & 0.99$\pm$0.04 & 0.78$\pm$0.03 & 10.94$\pm$0.21 & 2.15$\pm$0.54 &  \\
$L_\mathrm{peak, 0.3^{\prime\prime}}$ & ($10^{8}L_\odot$)  & 0.75$\pm$0.05 &  1.58$\pm$0.17 & 24.4$\pm$2.5  & $<$9.0 \\
$L_\mathrm{peak, 0.9^{\prime\prime}}$ & ($10^{8}L_\odot$)  & 1.39$\pm$0.09 &  3.46$\pm$0.37 & 63.2$\pm$6.4  & 22.1$\pm$7.1 \\
\hline
\multicolumn{6}{c}{kinematic properties} \\
\hline
$V_\mathrm{max}$ & (km s$^{-1}$) & 233$^{+24}_{-25}$ & 234$\pm24$ &  217$\pm22$ & &  \\
$\sigma_0$ & (km s$^{-1}$) & 94$\pm$9 & 94$\pm$10 & 77$\pm$8 & &  \\
$R_\mathrm{1/2, image}$ & (kpc) & 1.24$\pm$0.12 & 1.47$\pm$0.15 & 1.71$\pm$0.17 & &  \\
$R_\mathrm{1/2, visibility}$ & (kpc) & 1.24$\pm$0.13 & 1.53$\pm$0.18 & 2.01$\pm$0.08 &   \\
\hline
\end{tabular}
\end{center}
\end{table*}

\begin{figure*}[t]
\begin{center}
\includegraphics[scale=1.0]{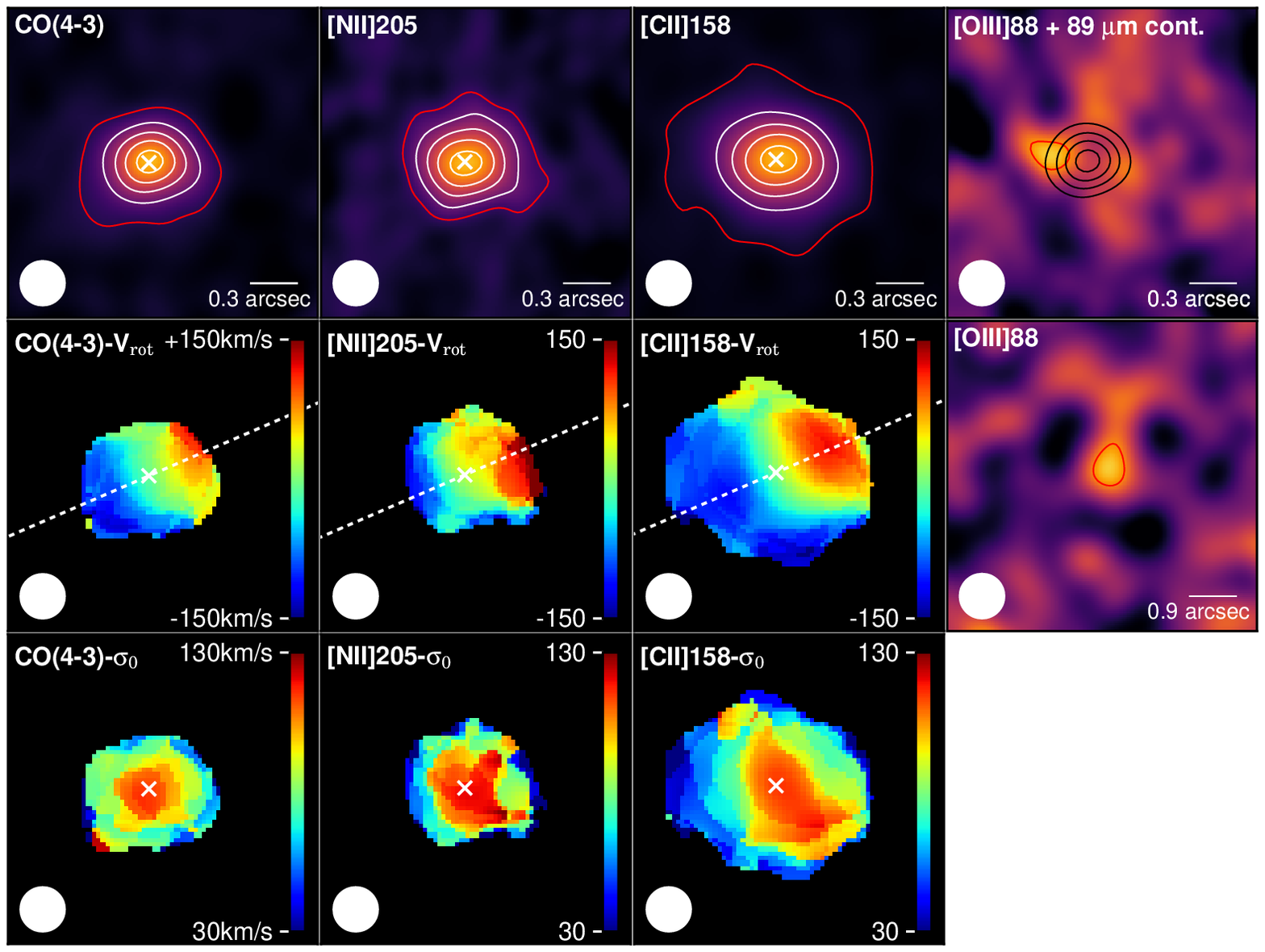}
\end{center}
\caption{
Top: From left to right, the 0.3\arcsec-resolution velocity-integrated flux maps of CO(4-3), [N~{\sc ii}], [C~{\sc ii}], \oiii line emission are displayed. 
The white contours denote the 90\%, 70\%, 50\%, 30\% values of the peak flux in CO(4-3), [N~{\sc ii}], [C~{\sc ii}] lines.
In the mostright panel, we overplot the same values of the rest-frame 89 $\mu$m continuum flux as black contours.
The red contours indicate the 3$\sigma$ level.
Middle: From left to right, the line-of-sight velocity maps of CO(4-3), [N~{\sc ii}], \cii emission and the 0.9\arcsec-resolution \oiii flux maps are shown.
The crosses and dashed white lines denote the kinematic center and the position angle of rotation.
Bottom: Local velocity dispersion maps of CO(4-3), [N~{\sc ii}], \cii emission.
}
\label{fig;map}
\end{figure*}

The far-infrared (FIR) fine structure lines of C, N and O offer valuable insights into the physical conditions in the interstellar medium (ISM) of SMGs at $z>3$.
The \cii158 $\mu$m line, primarily emitted from photodissociation regions (PDRs), is typically the brightest among the FIR fine structure lines \citep[e.g.,][]{2017ApJ...846...32D,2018ApJ...861...94H}.
Observations of [C~{\sc ii}], CO line and FIR continuum emission successfully characterize the physical properties (gas density and strength of radiation field) through theoretical models taking into account the chemistry, radiative transfer, and thermal balance of the neutral ISM \citep[e.g.,][]{1999ApJ...527..795K,1999RvMP...71..173H}.
The \nii205 $\mu$m line (or \nii122 $\mu$m) and the \oiii88 $\mu$m line (or \oiii52 $\mu$m) emission both arise only from ionized gas as the ionization potential is higher than that of hydrogen ($>13.6$ eV).
In SMGs at $z>3$, these lines have been detected with ground-based telescopes \citep[e.g.,][]{2010ApJ...714L.147F,2016ApJ...832..151P,2017ApJ...842L..16L,2018ApJ...856..174V,2018Natur.553...51M,2018ApJ...861...43P,2018ApJ...869L..22W} and with {\it Herschel} \citep[e.g.,][]{2011MNRAS.415.3473V,2018MNRAS.481...59Z}.
The line ratio of nitrogen and oxygen can be used as an indicator of gas-phase metallicity, which is commonly estimated from observations of rest-frame optical lines \citep{2017MNRAS.470.1218P,2018MNRAS.473...20R}.
The metallicity is one of the most important parameters to investigate galaxy formation because it imprints the past star formation histories.
For highly dust-obscured sources such as SMGs, the FIR lines have a big advantage over the optical lines in that they are less affected by dust extinction.

In this paper, we report results from Atacama Large Millimeter/submillimeter Array (ALMA) observations of an extreme starburst galaxy at $z=4.342$, COSMOS-AZTEC-1 hereafter, to study the spatial extent of the FIR fine structure lines and the physical conditions of gas in the PDRs and the ionized regions.
COSMOS-AZTEC-1 is one of the brightest SMGs, but not magnified by gravitational lensing \citep{2007ApJ...671.1531Y,2008MNRAS.385.2225S,2015MNRAS.454.3485Y, 2016ApJ...829L..10I}.
The high-resolution ALMA observations enables us to spatially-resolve the FIR fine structure lines and investigate the kinematic structures without uncertainties in lens modeling.

\section{Observations and results}

Following our previous observations of the CO(4-3) emission line with ALMA Band-3 receivers \citep{2018Natur.560..613T},
we have made new observations of the \nii205 $\mu$m, \cii158 $\mu$m, and \oiii88 $\mu$m lines with Band-6, 7, 9, respectively.
The observation date, the baseline length, the frequency coverage and the integration time are summarized in Table \ref{tab;obs}.
The data were calibrated in the standard manner using {\tt CASA} \citep{2007ASPC..376..127M}.
We create two cubes with different resolutions (0.3\arcsec and 0.9\arcsec), corresponding to 1 kpc and 3 kpc in radius, by applying a {\it uv} tapering and Gaussian smoothing (Table \ref{tab;obs}). 
We cleaned the cubes with a channel width of 50 km s$^{-1}$ down to the 1$\sigma$ level in a circular mask with a diameter of 1.5\arcsec. 
The resultant noise levels are listed in Table \ref{tab;obs}.
We also make a 0.3\arcsec-resolution map of the rest-frame 89 $\mu$m continuum emission by excluding the frequency range of the \oiii88 $\mu$m line.
The noise level is 0.43 mJy beam$^{-1}$.

For each emission line, we make 0.3\arcsec-resolution maps of velocity-integrated flux, velocity field and velocity dispersion in the same velocity range between --250 km s$^{-1}$ and +250 km s$^{-1}$ using the {\tt CASA/immoments} task. 
A 2$\sigma$ masking threshold was adopted for the velocity field and velocity dispersion maps. 
We detect the emission of CO(4-3), \nii and \cii line in the 0.3\arcsec-resolution maps at the level of 30$\sigma$, 22$\sigma$ and 56$\sigma$, respectively (Figure \ref{fig;map}).
For the \oiii emission, we see a 3$\sigma$ peak whose position is shifted from the peak of the dust emission by 0.2--0.3\arcsec in the 0.3\arcsec-resolution map.
The significance of the detection increases to 4.0$\sigma$ in the 0.9\arcsec-resolution map despite of the larger noise level (Figure \ref{fig;map}).
Given that the pixel at the position of COSMOS-AzTEC-1 has the maximum signal-to-noise ratio within the primary beam and the the maximum negative peak is identified at the level of -2.8$\sigma$, the \oiii detection should be real.
The 3$\sigma$ peak in the 0.3\arcsec-resolution map is likely produced by the fluctuations on the underlying component since interferometeric maps could create artificial clumps on an extended disk \citep{2016ApJ...833..103H,2018ApJ...859...12G}.
Therefore, we give the 3$\sigma$ upper limit in the 0.3\arcsec-resolution \oiii flux map and measure the peak fluxes in other maps ($S_\mathrm{peak}dv$ in Table \ref{tab;obs}).

\section{Analysis}

\subsection{Gas kinematics}\label{sec;kinematics}

For the CO(4-3), [N~{\sc ii}], [C~{\sc ii}] emission, the velocity field maps all show a monotonic gradient along the similar kinematic major axis, suggesting rotation of the gas (Figure \ref{fig;map}).
We fit the 0.3\arcsec-resolution cubes with dynamical models of a thick exponential disk using the {\tt GalPaK3D v1.9.1} code \citep{2015AJ....150...92B} to determine the maximum circular velocity $V_\mathrm{max}$, local velocity dispersion $\sigma_0$, half-light radius $R_{1/2}$, inclination and position angle.
We assume an arctan rotation curve and a constant $\sigma_0$ within a galaxy in the models.
From the modeling of the [C~{\sc ii}] cube with the highest signal-to-noise ratios, we derive that the inclination is 41.8$\pm0.2$ degree and the position angle is -65.7$\pm0.2$ degree.
The fitting errors are typically 2--3\%, based on the 95th percentile of the last 60\% of the Markov chain Monte Carlo (MCMC) chain for 20,000 iterations.
However, the comparison between data cubes with different clean parameters, leading to a different spatial resolution, shows systematic errors of $\sim$10\% in all parameters (Tadaki et al. in prep).
For fair comparisons of $V_\mathrm{max}$ and local velocity dispersion $\sigma_0$ among the lines,
we fix the inclination and the position angle to the \cii values for modeling of the CO and [N~{\sc ii}] cubes.
Table \ref{tab;obs} summarizes the best-fit values taking into account the systematic errors of 10\%.
The three lines have the similar $V_\mathrm{max}$ although they trace a different gas phase.
The agreement implies that the ionized gas feels the same gravitational potential as the associated PDR and molecular gas \citep{2018ApJ...854L..24U}.
The velocity dispersion of \cii emission, while its value is slightly smaller than that of CO and \nii emission, is in agreement with the other velocity dispersions given the level of precision in the measurements.
We confirm that COSMOS-AzTEC-1 is surely rotation-dominated with $V_\mathrm{max}/\sigma_0$=2.5--2.8.

\begin{figure}
\begin{center}
\includegraphics[scale=1.0]{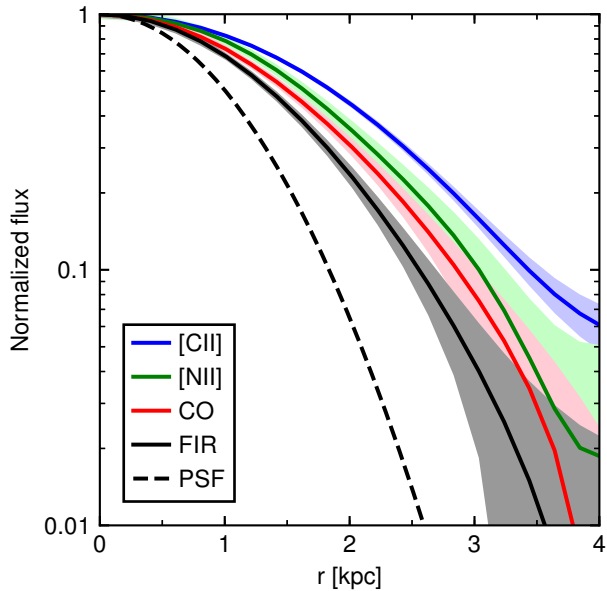}
\end{center}
\caption{
Radial profiles in the 0.3\arcsec-resolution flux maps of line and the rest-frame 89 $\mu$m continuum emission, labeled as FIR.
Shaded regions show the photometric uncertainties, not including the calibration errors.
A dashed line indicates a gaussian beam with FWHM=0.3\arcsec.
}
\label{fig;radial}
\end{figure}

The disk modeling above gives different $R_{1/2}$ among the lines as contrasted with the similar kinematic properties.
To verify this result, we fit the visibility data to exponential disk models using {\tt UVMULTIFIT} code \citep{2014A&A...563A.136M}.
The visibility-based analysis does not depend on clean parameters for reconstructing the images and is less affected by the spatial resolution.
The half-light radii are similar between the two methods (Table \ref{tab;obs}).
We also derive that the half-light radius of the rest-frame 89 $\mu$m continuum emission is $R_\mathrm{1/2,visibility}=$0.81$\pm$0.04 kpc.
The difference of the spatial extent is clearly seen in the radial profile of the surface brightness along the kinematic major axis (Figure \ref{fig;radial}).
The CO radial profile is similar to the \nii one although they have slightly different $R_{1/2}$.
The most conspicuous result is that the \cii emission is the most extended and the dust continuum is the most compact.
As the rest-frame 89 $\mu$m is generally close to the peak wavelength of dust emission heated by star formation, 
the continuum emission directly traces the FIR luminosities and thus dust-obscured star formation rates.
This significant difference between \cii and FIR would lead to a radial variation in the strength of radiation field, which is seen in both nearby \citep[e.g.,][]{2017ApJ...842..128K, 2017ApJ...846...32D} and high-redshift galaxies \citep{2018ApJ...867..140L}.

\subsection{Far-infrared luminosities}

\begin{figure}
\begin{center}
\includegraphics[scale=1.0]{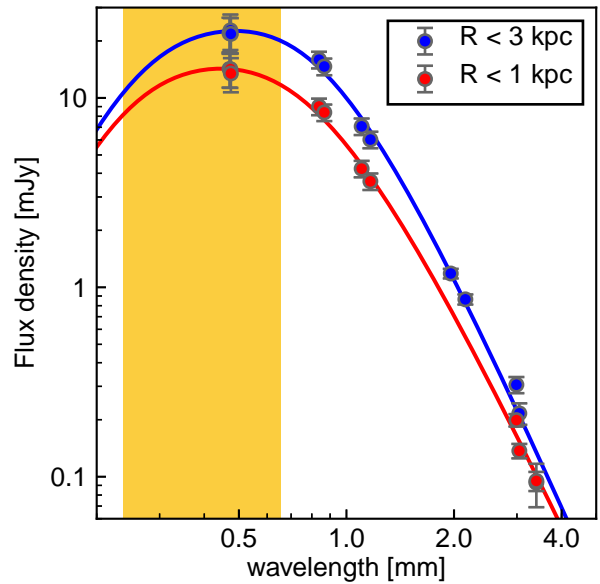}
\end{center}
\caption{
Dust continuum SEDs with the best-fit modified blackbody radiation models with a dust temperature of $T_\mathrm{dust}=54^{+8}_{-3}$ K (blue line) and $T_\mathrm{dust}=59^{+5}_{-7}$ K (red line). 
The blue and red circles indicate the continuum fluxes measured in the central 3 kpc and 1 kpc, respectively.
We compute the far-infrared luminosities of the best-fit models in the rest-frame wavelength range of 42.5 $\mu$m--122.5 $\mu$m shown by a yellow shaded region.
}
\label{fig;sed}
\end{figure}

ALMA multi-band observations provide multi-data point of continuum emission, constraining a spectral energy distribution (SED) of dust component.
We create 0.3\arcsec- and 0.9\arcsec-resolution continuum maps in each spectral window for the Band-3, 6, 7, 9 data and also use 0.9\arcsec-resolution Band-4 data (2 mm; \citealt{2018Natur.560..613T}).
The flux uncertainties are mainly dominated by the flux calibration errors (5\% at Band-3,4, 10\% at Band-6,7, 20\% at Band-9; ALMA Technical Handbook) rather than the signal-to-noise ratios of the detections.
The measured continuum fluxes are given in Table \ref{tab;continuum}.
To estimate the far-infrared luminosities $L_\mathrm{FIR}$ in the rest-frame wavelength range of 42.5 $\mu$m--122.5 $\mu$m, we model the observed SEDs at 10 bands in the central 1 kpc region (0.3\arcsec-resolution map) and at 12-bands in the central 3 kpc region (0.9\arcsec-resolution map) using the {\tt CIGALE} code \citep{2005MNRAS.360.1413B,2018arXiv181103094B}.
We adopt a simple analytic model with a single modified black body radiation, characterized by dust temperature $T_\mathrm{dust}$ and an emissivity index $\beta$, and a power-low emission \citep{2012MNRAS.425.3094C}.
The power-low component has little contribution to our modeling since short wavelength data is not included ($\lambda<80~\mu$m in the rest-frame).
Figure \ref{fig;sed} shows the observed SEDs and the best-fit models giving $L_\mathrm{FIR}=5.3\pm1.1\times10^{12}~L_\odot$, $T_\mathrm{dust}=59^{+5}_{-7}$ K and $\beta=2.1^{+0.2}_{-0.1}$ in the central 1 kpc region and $L_\mathrm{FIR}=7.9\pm1.7\times10^{12}~L_\odot$, $T_\mathrm{dust}=54^{+8}_{-3}$ K and $\beta=2.4^{+0.1}_{-0.2}$ in the central 3 kpc region.
The central 1 kpc region has a slightly higher dust temperature than the outer regions but the difference is within the fitting errors.

\begin{table}[!h]
\caption{Dust continuum fluxes in AzTEC-1. \label{tab;continuum}}
\begin{center}
\begin{tabular}{lcc}
\hline
wavelength & $S_\mathrm{peak, 0.9^{\prime\prime}}$ & $S_\mathrm{peak, 0.3^{\prime\prime}}$ \\
($\mu$m) & (mJy beam$^{-1}$) &(mJy beam$^{-1}$) \\
\hline
471 & 21.97 $\pm$ 4.55 & 14.12 $\pm$ 2.87 \\ 
473 &  22.74 $\pm$ 4.83 & 14.11 $\pm$ 2.88 \\ 
475 &  21.82 $\pm$ 4.61 & 13.31 $\pm$ 2.73 \\ 
838 &  15.91 $\pm$ 1.61 & 9.02 $\pm$ 0.91 \\ 
867 &  14.64 $\pm$ 1.48 & 8.41 $\pm$ 0.84 \\ 
1103 &  7.08 $\pm$ 0.71 & 4.23 $\pm$ 0.42 \\ 
1167 &  6.03 $\pm$ 0.61 & 3.63 $\pm$ 0.36 \\ 
1958 &  1.18 $\pm$ 0.07 & -- \\ 
2152 &  0.86 $\pm$ 0.05 & -- \\ 
2989 &  0.31 $\pm$ 0.03 & 0.20 $\pm$ 0.01 \\ 
3046 &  0.22 $\pm$ 0.03 & 0.14 $\pm$ 0.01 \\ 
3400 &  0.09 $\pm$ 0.02 & 0.10  $\pm$ 0.01 \\ 
\hline
\end{tabular}
\end{center}
\end{table}

\subsection{Gas properties in PDR}\label{sec;pdr}

The \cii emission is more extended than the CO(4-3) and the rest-frame 89 $\mu$m continuum emission (Section \ref{sec;kinematics}). 
This does not necessarily mean different beam filling factors among the three emission because the kpc-scale resolution is much larger than individual PDRs.
Given that we observe the cumulative emission from many PDRs,
the difference in the spatial distributions would reflect a radial variation in the typical gas properties.
We calculate the luminosities of CO and \cii emission in both the central 1 kpc region and the central 3 kpc region, with taking into account the systematic errors on the flux calibration as well as the random errors based on the signal-to-noise ratio ($L_\mathrm{peak}$ in Table \ref{tab;obs}).
By comparing our measurements with values predicted by theoretical models of \cite{1999ApJ...527..795K}, we determine the hydrogen gas density $n_\mathrm{H}^\mathrm{PDR}$ and the strength of the incident far-ultraviolet (FUV) radiation fields with $6<h\nu<13.6$ eV, $G_0$.

Standard models of \cite{1999ApJ...527..795K} consider a simple geometry of one-dimensional plane-parallel slabs illuminated from one side by a FUV flux $G_0$.
\cii line and dust continuum emission are generally optically thin while CO emission is optically thick.
Therefore, we increase the observed CO luminosities by a factor of two to count the emission from the far side.
As \cii emission comes from ionized regions as well as PDRs, we need to subtract the contribution of ionized gas from observed \cii luminosities.
A \nii 205 $\mu$m line is useful for estimating \cii luminosities arising from ionized regions [C~{\sc ii}]$^\mathrm{ion}$ since its critical density and excitation energy are similar to those of a \cii line.
Here, we assume a line ratio of [C~{\sc ii}]$^\mathrm{ion}$/[N~{\sc ii}]=2, predicted from photo-ionization models, in the ionized gas (Section \ref{sec;metallicity}).
The fraction of \cii originating from PDRs is [C~{\sc ii}]$^\mathrm{PDR}$/[C~{\sc ii}]=81\% in the central 1 kpc region and [C~{\sc ii}]$^\mathrm{PDR}$/[C~{\sc ii}]=84\% in the central 3 kpc region, which are similar to the typical values in local luminous and ultra luminous infrared galaxies (LIRGs and ULIRGs) \citep{2017ApJ...846...32D}. 

\begin{figure}
\begin{center}
\includegraphics[scale=1.0]{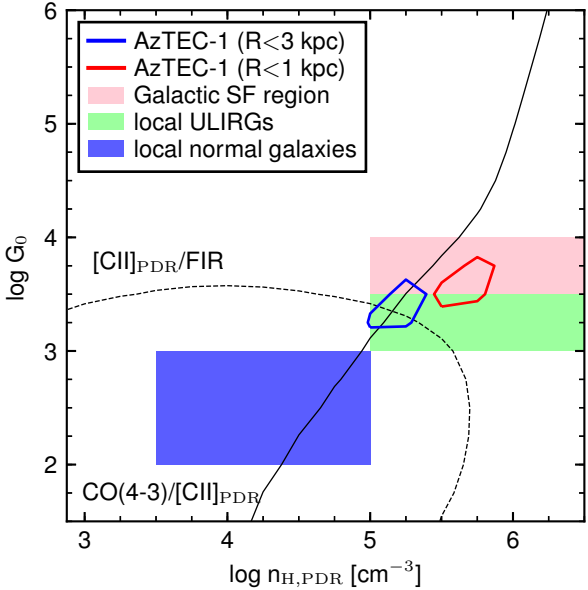}
\end{center}
\caption{
$\chi^2$ distributions, based on PDR modeling and our observations, as a function of hydrogen density and FUV radiation field.
The blue and red contours correspond to the 68\% confidence levels ($\Delta \chi^2=2.3$) in the central 3 kpc and 1 kpc region of AzTEC-1, respectively.
The solid and dashed lines show the CO(4-3)/[C~{\sc ii}]$_\mathrm{PDR}$ luminosity ratio and [C~{\sc ii}]$_\mathrm{PDR}$/FIR ratio measured in the central 3 kpc region, respectively.
We overplot the typical ranges for Galactic OB star formation region, local ULIRGs and local normal galaxies by the red, green and blue shaded regions \citep{2010ApJ...724..957S}.
}
\label{fig;pdr}
\end{figure}

With our measurements and the predictions by the PDR model, we compute chi-square values for two luminosity ratios of [C~{\sc ii}]$^\mathrm{PDR}$/FIR and [C~{\sc ii}]$^\mathrm{PDR}$/CO in the ranges of $n_\mathrm{H}^\mathrm{PDR}=10^{3-7}$ cm$^{-3}$ and $G_0=10^{1-6}$.
We derive the appropriate parameters of $n_\mathrm{H}^\mathrm{PDR}=10^{5.5-5.75}$ cm$^{-3}$ and $G_0=10^{3.5-3.75}$ in the central 1 kpc region and $n_\mathrm{H}^\mathrm{PDR}=10^{5.0-5.25}$ cm$^{-3}$ and $G_0=10^{3.25-3.5}$ in the central 3 kpc region with the confidence level of 68\% (Figure \ref{fig;pdr}).
The central 1 kpc region likely has a higher gas density compared to the outer region, which is expected given the strong concentration of FIR emission.
The physical conditions are close to those found in Galactic OB star formation regions and local ULIRGs and are also consistent with previous studies for other high-redshift SMGs \citep[e.g.,][]{2010ApJ...714L.162H,2010ApJ...724..957S,2019arXiv190110027R}.
Some studies using dense gas tracers also support a high-density gas in SMGs \citep[e.g.,][]{2011MNRAS.410.1687D,2014ApJ...785..149S,2017ApJ...850..170O}.

\begin{figure}
\begin{center}
\includegraphics[scale=1.0]{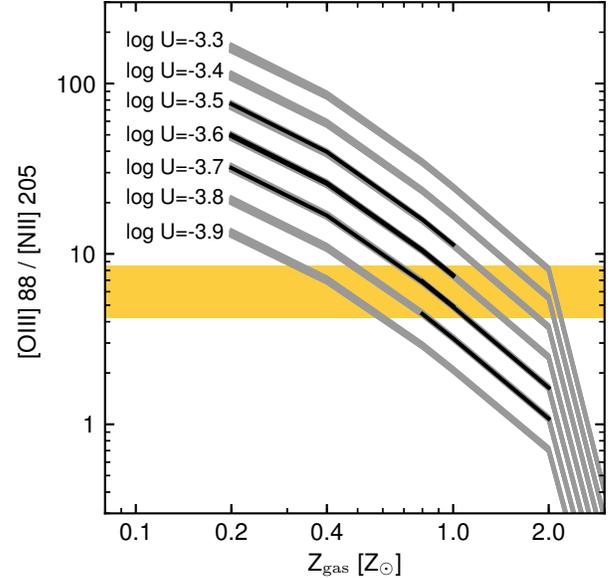}
\end{center}
\caption{
[O~{\sc iii}]/[N~{\sc ii}] luminosity ratio as a function of gas-phase metallicity.
The yellow shaded region shows our measurement in the central 3 kpc region of AzTEC-1, $L_\mathrm{[OIII]}/L_\mathrm{[NII]}=6.4\pm2.2$.
We plot the line ratios of the fiducial photoionization models, that are consistent with our PDR modeling, by black lines and those of all explored models by gray lines.
}
\label{fig;oiiinii}
\end{figure}

\subsection{Gas-phase metallicity}\label{sec;metallicity}

We have successfully measured the line ratio of [O~{\sc iii}] 88 $\mu$m/[N~{\sc ii}] 205 $\mu$m in the central 3 kpc region of COSMOS-AzTEC-1 at $z=4.3$, which is the highest redshift where both nitrogen and oxygen lines are detected. 
We interpret the ratio as an indicator of gas-phase metallicity $Z_\mathrm{gas}$.
However, [O~{\sc iii}]/[N~{\sc ii}] ratios strongly depend on the hydrogen density in ionized regions $n_\mathrm{H}^\mathrm{ion}$ and the dimentionless ionization parameter $U^\mathrm{ion}$, defined as $U^\mathrm{ion}=\phi_\mathrm{H}/n_\mathrm{H}^\mathrm{ion}c$ where $\phi_\mathrm{H}$ is the flux of ionizing photons with $h\nu>13.6$ eV and $c$ is the speed of light.
In the previous sections, we derived the hydrogen density in PDRs $n_\mathrm{H}^\mathrm{PDR}$ and the flux of FUV radiation $G_0$.
These two parameters are closely related to $n_\mathrm{H}^\mathrm{ion}$ and $\phi_\mathrm{H}$.
We use the spectral synthesis code {\tt Cloudy v17.01} \citep{2017RMxAA..53..385F} to calculate $n_\mathrm{H}^\mathrm{PDR}$ and $G_0$ as well as the predicted [O~{\sc iii}]/[N~{\sc ii}] ratio as functions of $n_\mathrm{H}^\mathrm{ion}$, $U^\mathrm{ion}$ and $Z_\mathrm{gas}$.
We generate the input spectra of a constant star formation model with an age of 1 Myr by using Binary Population and Spectral Synthesis ({\tt BPASS v2.0}) code \citep{2016MNRAS.462.3302E,2016MNRAS.456..485S}.
We also assume that the stellar metallicity is lower than gas-phase metallicity by a factor of 5 since the duration of the extreme starburst is likely shorter than a timescale for metal enrichment by type Ia supernovae ($\sim1$ Gyr).
Adopting models with massive star binaries and lower stellar metallicities is motivated by recent results in star-forming galaxies at $z\sim2$ \citep{2016ApJ...826..159S}.
Gas element abundance patterns and other parameters in the Cloudy run are the same as the previous calculations by \cite{2011A&A...526A.149N}.

We find that only models with $n_\mathrm{H}^\mathrm{ion}=10^{4.2}-10^{4.3}$ cm$^{-3}$ and $U^\mathrm{ion}=10^{-3.8}-10^{-3.5}$ satisfy the physical conditions ($n_\mathrm{H}^\mathrm{PDR}=10^{5.0-5.25}$ cm$^{-3}$ and $G_0=10^{3.25-3.5}$) constrained by our PDR modeling.
Under the restrictions of $n_\mathrm{H}^\mathrm{ion}$ and $U^\mathrm{ion}$, [O~{\sc iii}]/[N~{\sc ii}] ratios primarily depend on the gas-phase metallicity (Figure \ref{fig;oiiinii}). 
The measured luminosity ratio of $L_\mathrm{[OIII]}/L_\mathrm{[NII]}=6.4\pm2.2$ corresponds to a near solar metallicity, $Z_\mathrm{gas}=0.7-1.0~Z_\odot$ in the fiducial model with $U^\mathrm{ion}=10^{-3.7}$.
Given the stellar mass of $M_\star\sim10^{11}~M_\odot$ in COSMOS-AzTEC-1 \citep{2018Natur.560..613T}, the metallicity is by a factor of 2--3 lower than that of similarly massive galaxies at $z\sim0$ based on observations of rest-optical nebular lines \citep{2008ApJ...681.1183K}.
At a fixed stellar mass, the metallicity of galaxies decreases as a function of increasing redshift.
\cite{2016ApJ...822...42O} have measured the metallicities with optical lines for star-forming galaxies at $z=3-4$ in the main stellar mass range of $M_\star=10^{9.5}-10^{10.5}~M_\odot$, showing a positive correlation between stellar mass and metallicity (see also \citealt{2008A&A...488..463M}).
The extrapolation of the mass--metallicity relation at $z=3-4$ gives $Z_\mathrm{gas}\sim1.0~Z_\odot$ for massive galaxies with $M_\star=10^{11}~M_\odot$, which is consistent with our measurement.

\section{Discussion}

We have reported the physical properties of the CNO fine structure lines (\cii158 $\mu$m, \nii205 $\mu$m, and \oiii88 $\mu$m) in a bright unlensed submillimeter galaxy at $z=4.3$, which is the highest redshift where both nitrogen and oxygen lines are detected.
Our deep and high-resolution data show that the ionized, PDR and molecular gas have the similar kinematic properties.
We confirm that COSMOS-AzTEC-1 is rotation-dominated with $V_\mathrm{max}/\sigma_0$=2.5--2.8, which is close to the values for massive quiescent galaxies at $z\sim2$ \citep[e.g.,][]{2017Natur.546..510T,2018ApJ...862..126N}.
Since these galaxy populations are likely the progenitors of the most massive slow rotators at $z=0$ \citep[e.g.,][]{2016ARA&A..54..597C}, they have to lose significant angular momentum in the intervening time \citep{2017ApJ...841L..25T}.
Cosmological simulations predict that {\it dry} mergers efficiently spin down galaxies while {\it wet} mergers increase the angular momentum \citep[e.g.,][]{2014MNRAS.444.3357N, 2018MNRAS.473.4956L}.
The high-redshift progenitors of slow rotators would change the kinematic properties at $z<2$.

We have also determined the physical conditions of gas in PDRs and ionized regions with the CNO emission.
Our PDR modeling indicates that most of the \cii emission arises from dense PDRs with $n_\mathrm{H}^\mathrm{PDR}=10^{5.0-5.5}$ cm$^{-3}$ and $G_0=10^{3.25-3.75}$, which are likely associated with massive star formation.
These PDR parameters constrain the gas density in ionized regions and the ionizing flux, leading to the ionization parameters.
Using the fiducial photoionization models and the measured [O~{\sc iii}]/\nii ratios, we find COSMOS-AzTEC-1 to be a chemically-evolved system with $Z_\mathrm{gas}=0.7-1.0~Z_\odot$, which are consistent with previous studies of other dusty star-forming galaxies at high-redshift \citep{2012A&A...542L..34N,2018MNRAS.473...20R}.
The 0.3\arcsec-resolution \oiii map also gives the upper limit of $L_\mathrm{[OIII]}/L_\mathrm{[NII]}<5.7$ in the central 1 kpc region, which is not much larger than the value in the central 3 kpc region, $L_\mathrm{[OIII]}/L_\mathrm{[NII]}=6.5\pm2.2$.
Provided that the ionization parameters are constant within the galaxy, this result would reject positive radial gradients with lower metallicity in the center as sometimes seen in star-forming galaxies at $z=2-4$ \citep[e.g.,][]{2010Natur.467..811C, 2013ApJ...765...48J}.

We note that our metallicity measurements are based on the assumption that ionizing sources (OB stars) are all associated with dense PDRs, motivated by the high gas density and the intense radiation field.
In another extreme case that OB stars are randomly distributed with respect to the PDRs \citep{1990ApJ...358..116W}, 
the average ionized gas density is not necessarily connected to the gas density in individual PDR clouds.
Measurements of \nii 122 $\mu$m/\nii 205 $\mu$m or [O~{\sc iii}]52/[O~{\sc iii}]88 line flux ratios allow us to directly estimate the gas density in the ionized region without the assumption of geometry.
In nearby star-forming galaxies, the ionized gas density is estimated to be $n_{H}^\mathrm{ion}=10^{1-2}$ cm$^{-3}$ \citep[e.g.,][]{2016ApJ...826..175H, 2017ApJ...846...32D}, which is much lower than expected in COSMOS-AzTEC-1.
The high gas density is preferred in extreme starburst galaxies, but it needs to be verified by direct measurements based on \nii 122 $\mu$m/\nii 205 $\mu$m or [O~{\sc iii}]52/[O~{\sc iii}]88 ratios.
Ionization parameters are also another factor leading to the uncertainties of metallicity measurements.
\cite{2011A&A...526A.149N} find that the [O~{\sc iii}]88 $\mu$m/[N~{\sc iii}] 57 $\mu$m ratio is a good tracer of the gas-phase metallicity, being weakly dependent on ionization parameters.
For galaxies at $z<4.6$, [N~{\sc iii}] 57 $\mu$m emission are not, unfortunately, accessible with ALMA.
Comparing the [N~{\sc ii}] and [O~{\sc iii}] lines with the radio free-free continuum would be a good test for verifying our approach through PDR modeling since it is an independent method to measure the metallicity \citep{1981ApJ...250..186H,2015ApJ...806..260F, 2018ApJ...867..140L}.
In spite of these uncertainties, observations of CNO emission in SMGs would open a new avenue for understanding chemical evolution of massive galaxies in early universe.

\ 

We thank the referee for valuable suggestions that improved the paper.
We also thank the ALMA staff and in particular the EA-ARC staff for their support.
This paper makes use of the following ALMA data: ADS/JAO.ALMA\#2017.1.00300.S, 2018.1.00081.S, 2017.1.00127.S. ALMA is a partnership of ESO (representing its member states), NSF (USA) and NINS (Japan), together with NRC (Canada) and NSC and ASIAA (Taiwan) and KASI (Republic of Korea), in cooperation with the Republic of Chile. The Joint ALMA Observatory is operated by ESO, AUI/NRAO and NAOJ.
K.T. acknowledges support by Grant-in-Aid for JSPS Research Fellow JP17J04449. 
Data analysis was in part carried out on the common-use data analysis computer system at the Astronomy Data Center (ADC) of the National Astronomical Observatory of Japan.

\end{document}